\documentclass[12pt,preprint]{aastex}

\shorttitle{Photospheric and Circumstellar Asymmetries}
\shortauthors{Ragland et al.}

\begin{document}


\title{First Surface-resolved Results with the IOTA Imaging Interferometer: 
Detection of Asymmetries in AGB stars}


\author{Ragland, S.\altaffilmark{1,2}, 
Traub, W.A.\altaffilmark{3,1},
Berger, J.-P.\altaffilmark{4},
Danchi, W.C.\altaffilmark{9},
Monnier, J. D.\altaffilmark{6},
Willson, L. A.\altaffilmark{11},   
Carleton, N. P.\altaffilmark{1},
Lacasse, M. G.\altaffilmark{1},
Millan-Gabet, R.\altaffilmark{5},
Pedretti, E.\altaffilmark{6},
Schloerb, F. P.\altaffilmark{7},
Cotton, W. D.\altaffilmark{8}, 
Townes, C.H.\altaffilmark{10}, 
Brewer, M.\altaffilmark{7}, 
Haguenauer, P.\altaffilmark{12}, 
Kern, P.\altaffilmark{4}, 
Labeye, P.\altaffilmark{13}, 
Malbet, F.\altaffilmark{4}, 
Malin, D.\altaffilmark{7},
Pearlman, M.\altaffilmark{1},
Perraut, K.\altaffilmark{7}, 
Souccar, K.\altaffilmark{7}, 
Wallace, G.\altaffilmark{7}}

\altaffiltext{1}{Harvard-Smithsonian Center for Astrophysics, 60 Garden 
Street, Cambridge, MA 02138}
\altaffiltext{2}{Presently at California Association for Research in Astronomy, 65-1120 Mamalahoa Hwy, Kamuela, HI 96743; sragland@keck.hawaii.edu}
\altaffiltext{3}{Jet Propulsion Laboratory,  M/S 301-451, 4800 Oak Grove Dr., Pasadena CA, 91109}
\altaffiltext{4}{Laboratoire d'Astrophysique de Grenoble, 414 Rue de la 
Piscine, F-38400 Saint Martin d'Heres, France.}
\altaffiltext{5}{Michelson Science Center, California Institute of 
Technology, Pasadena, 770 S. Wilson Ave., Pasadena, CA 91125}
\altaffiltext{6}{University of Michigan at Ann Arbor, Department of 
Astronomy, 500 Church Street, Ann Arbor, MI 48109-1090.}
\altaffiltext{7}{University of Massachusetts at Amherst, Department of 
Astronomy, LGRT-B 619E, 710 North Pleasant Street, Amherst, MA 01003-9305.}
\altaffiltext{8}{National Radio Astronomy Observatory, 520 Edgemont Road, 
Charlottesville, VA 22903}
\altaffiltext{9}{NASA Goddard Space Flight Center, Exoplanets \& Stellar 
Astrophysics, Code 667, Greenbelt, MD 20771}
\altaffiltext{10}{University of California at Berkeley, Space Science 
Laboratory, Berkeley, CA 94725-7450}
\altaffiltext{11}{Department of Physics and Astronomy, Iowa State University,
Ames IA  50014}
\altaffiltext{12}{ALCATEL Space Industries, 100 boulevard du Midi,
BP99, 06322 Cannes, France}
\altaffiltext{13}{LETI, CEA-Grenoble, 17 rue des Martyrs, 
38 054 Grenoble CEDEX 9, France}





\begin{abstract}
We have measured non-zero closure phases for 
about 29\% of our sample of 56 nearby Asymptotic Giant Branch (AGB) stars, 
using the 3-telescope 
Infrared Optical Telescope Array (IOTA) interferometer at near-infrared 
wavelengths (H-band) and with 
angular resolutions in the range 5-10 milliarcseconds. These non-zero 
closure phases can only be generated by asymmetric brightness distributions
of the target stars or their surroundings.
We discuss how these results were obtained, and how they might be interpreted in
terms of structures on or near the target stars. 
We also report measured angular sizes 
and hypothesize that most Mira stars would show detectable asymmetry if observed with adequate angular resolution.
 
\end{abstract}




\section{Introduction}

The stars in this study are all AGB stars, that is, stars found at or near
the tip of the Asymptotic Giant Branch in the HR diagram. They are low to 
intermediate mass stars, having already spent most of their lives as normal 
stars, and 
currently heading towards their deaths probably in the form of planetary 
nebulae, leaving the central star as a white dwarf. Most AGB stars are 
variable in brightness; those with relatively regular and large amplitude
visual variations ($>$ 2.5 mag) with periods in the range 100 - 1000 days 
are classified as Mira variables. The Miras 
and some of the other, semi-regular (SR) or irregular (Irr) variables have
observed mass loss rates ranging from 10$^{-7}$ to $>$ 10$^{-5}$ M$_\odot$
per year \citep{Knapp85}. Diameter changes, opacity changes, and possibly other 
processes such as
convection contribute to the brightness variation in these stars.

The Mira stage of evolution has been identified 
as marking the onset of the ``superwind'' phase, 
i.e. that evolutionary stage where mass loss rates  
rapidly increase and result in the 
termination of AGB evolution (Bowen \& Willson 
1991, Willson 2000).  These stars thus serve as 
markers for the tip of the AGB in various 
populations, something already known for the 
shorter period cases from the few Miras that 
appear in globular clusters such as 47 Tuc 
(Frogel, Persson \& Cohen 1981). 

Miras with close 
companion white dwarfs usually are classified as 
symbiotic systems (Allen 1984;  Whitelock 1987; 
Luthardt 1992; Belczy{\' n}ski et al. 2000).   A few 
Miras are known to have companions but are not 
(or are only very mildly) symbiotic systems; this 
includes {\it o} Cet = Mira, with a probable WD 
companion in a multi-century orbit (Reimers \& 
Cassatella 1985; Wood \& Karovska 2004). 
Statistics for the binarity of Miras are otherwise 
quite uncertain, in part because the expected 
orbital velocity amplitudes for a close companion, 
30 km/s at 1 AU and 10 km/s around 5 AU, are very 
similar to the shock amplitudes of 20-30 km/s 
produced by the Mira pulsation itself (Hinkle, Scharlach \& Hall 1984).

In this paper we use the word ``asymmetry'' to mean that part of the 
2-dimensional brightness distribution which cannot be made symmetric with 
respect to a reflection through a point. Thus, for example, an elliptical 
uniform disc or an equal-brightness binary system are both symmetric, but 
a binary with unequal brightness or a star with an off-centered bright/dark spot 
is asymmetric.

Departure from circular symmetry has been known in AGB stars from various 
high angular resolution observations
\citep{Karovska91, Wilson92, Haniff92, Richichi95, Ragland96, Weigelt96, 
Karovska97, Tuthill97, Lattanzi97, Wittkowski98, Tuthill99, 
Tuthill00, Hofmann00, Thompson02, Monnier04a, Weiner06}.
The observed departures from circular symmetry have been interpreted 
either in terms of elliptical distortions or an otherwise symmetric
photosphere containing localized compact features. 
However, no consensus exists as to the mechanism that would cause such 
departures from apparent circular symmetry.

Dust shells surrounding AGB stars have observed asymmetries as well such as in 
Mira \citep{Lopez97}, the carbon stars IRC+10216 (CW Leo; \citet{Tuthill00}), 
CIT 6 \citep{Monnier00} and IK Tau \citep{Weiner06} among others. The connection
between apparent surface features and the morphology of the dust shells has not been 
established.  

About 50\% of all planetary nebulae (PN) display bipolar symmetry 
\citep{Zuckerman86}, but only a small fraction of circumstellar 
envelopes show bipolarity. A surprisingly large number of proto-PN show 
roughly 
circular arcs surrounding a bipolar core, suggesting that in most cases 
the AGB mass loss is spherically symmetric and the asymmetry 
seen in the PN occurs well after the Mira stage (Su 2004; 
Willson \& Kim 2004).
Recent studies of jets around a few
AGB stars \citep{Kellogg01, Imai02, Sahai03, Sokoloski03, Brocksopp04} 
from radio, x-ray or Hubble Space Telescope (HST) observations suggest 
that those stars showing substantial asymmetry may all have a low mass 
stellar companion accreting mass from the AGB primary.
Recent SiO maser observations of AGB stars show departures from spherical 
symmetry (\citet{Diamond94}, \citet{Green95}, \citet{Diamond03}, \citet{Cotton04}, \citet{Soria04}). 
The observed circumstellar SiO masers tend to occur in 
clumpy, partial rings centered on the central star (\citet{Diamond94}).
Cotton et al. (2002) observed 9 stars in SiO, at least 
two of them known binaries, and used the modeling 
of Humphreys et al. (2002) in discussing the 
results. 
SiO maser emission 
comes from $\sim$ 2AU or $\sim$ 2R$_*$ where 
the outflow velocity gradient along the line of sight is small. The special conditions 
required for maser emission potentially gives rise to bias in the statistics 
of asymmetry in the sample population of stars. 

In this paper we report the initial results from one phase of a larger 
program, the Mira Imaging Project, to investigate asymmetries in AGB stars 
using three interferometer
facilities, each capable of making closure phase measurements. These 
facilities are the IOTA, the Infrared Spatial Interferometer (ISI), and 
the Very Long Baseline Array (VLBA). 
The present paper focuses on IOTA results.

Subsequent to the work reported here, and as a part of the ongoing Mira Imaging Project, selected Mira targets with positive closure phase
signal from our survey have been re-visited at different pulsational phases, baselines, position angles, 
and wavelengths in order to characterize the observed asymmetry. The results of this ongoing 
study will be presented elsewhere. In this article, we present the initial survey results for
all our targets. 

\section{Observations}

The observations reported here were carried out during the commissioning 
phase of the IOTA 3-telescope array \citep{Traub04} and 
integrated-optics beam-combiner, IONIC \citep{Berger04}, operating 
in the H-band (1.65 $\mu$m) atmospheric window.  
Observations of binary stars taken with the same instrumental configurations were  
reported by \citet{Monnier04b} and \citet{Kraus04}. 
We report here the results of the first phase of our program in which we have studied 56 evolved giants 
(Tables~\ref{result} \& \ref{negResult}) including 35 Mira stars, 18 SR variables and 
3 Irr variables\footnotemark[1]\footnotetext[1]{$\delta$2 Lyr is classified in the Combined General 
Catalogue of Variable Stars (CGCVS; \citet{Samus04}) with an 
uncertainty as a SR variable, and no period estimation is available in 
the literature. We consider this target as being an Irr variable for the 
purpose of this paper.} looking for asymmetry in their brightness profiles.

We report observations taken during six observing runs during May 2002 to May 2003. Observations were taken either with a standard H band filter 
($\lambda_o$ = 1.65 $\mu$m, $\Delta \lambda$ = 0.3 $\mu$m) or with a narrow-band filter
($\lambda_o$ = 1.64 $\mu$m, $\Delta \lambda$ = 0.1 $\mu$m). Typically, five minutes of program star 
observations were followed by nearby calibrator observations under 
identical instrumental configurations. For the observations taken during March 2003 and May 2003 observing runs, we used ND filters for the bright targets since we had excellent optical throughput with newly coated primary mirrors and well optimized beam-train. On each star, we record 4 sets of data
files each containing about 500 scans. A scan consists of changing the optical path difference between two beams by roughly 75 $\mu$m in saw tooth form. We then take about 400 scans of shutter data for calibration. The shutter data sequence consists of allowing only one beam at a time (telescope A, B and then C) and at the end blocking all three beams. Each scan takes about 100 ms.  


  
All targets were observed with a three-baseline interferometer configuration, forming 
a closed triangle. Earth rotation enables closure-phase
measurements at slightly different projected baselines (and hence different 
closed triangles) when observations are made at different hour angles. 
We have adopted baseline bootstrapping \citep{Mozurkewich92} at the IOTA 
whereby fringes are tracked on two short baselines, while 
science data are recorded on all three baselines simultaneously, enabling 
low visibility measurements on the third (long) baseline. 
Details of the detector camera and the fringe tracker algorithm used for this work 
are reported by \citet{Pedretti04} and  \citet{Pedretti05} respectively. 

IOTA's maximum baseline of B=38 m yields an angular resolution of 
$\lambda$/2B $\simeq$ 4 mas at 1.65 $\mu$m. The present faint limit with the
IONIC beam-combiner is H $\simeq$ 7 for the broad band filter and 
H $\simeq$ 5 for the three narrow band filters. For the 
difficult case of observing well resolved Mira stars with the H filter at 
or below 5\% visibility level, the limiting magnitude is H $\simeq$ 4. 
The limiting magnitude of the star tracker at IOTA, for these observations,
was V $\simeq$ 12 for late-type stars. The angular resolution of 
our short south-east arm at IOTA is lower than
that of the North-east arm, meaning that we could miss some asymmetry if it 
were predominantly parallel to the projected south-east baseline of the interferometer. 

Pulsation periods for all Mira and SR variables (53 out of 56 program 
stars) are from Combined General Catalog of Variable Stars (CGCVS). Among these 53 program stars, 37 also have 
period estimations from American Association of Variable Star Observers 
(AAVSO) derived using a data window centered on JD 2452000 
\citep{Templeton04} enabling us to validate the CGCVS periods. 
In addition, AAVSO has tentative or very tentative periods for three 
more SR stars. The CGCVS periods are consistent with available AAVSO periods
for all but four of our program stars. Interestingly, the AAVSO periods for 
all four discrepant stars, namely X Cnc, BG Ser, UU Aur and W Ori, 
are roughly twice that of CGCVS periods although the AAVSO periods
for two of them, namely UU Aur and W Ori, are either tentative or 
very tentative values. For completeness, one of the Irr variable, namely, 
TX PSc has a very tentative AAVSO period of 255.5 days.  

\section{Data Reduction}

The recorded interferograms were reduced with an IDL code package developed
by one of us (Ragland). 
Our single-mode integrated-optics beam-combiner chip \citep{Berger04} has 3 input beams ($I_a$, $I_b$, $I_c$), and
6 output beams ($I_i$, $i = 1-6$). Each input is split into 2 parts and 
coupled to the outputs as follows: ($a,b$) $\leftrightarrow$ ($1,2$); 
($a,c$) $\leftrightarrow$ ($3,4$); ($b,c$) $\leftrightarrow$ ($5,6$).
The complimentary outputs ($I_2$, $I_4$ \& $I_6$) have the same information 
as the normal outputs ($I_1$, $I_3$ \& $I_5$) except for a $\pi$ fringe intensity phase shift. 
Hence the normal and the complimentary outputs could be combined in order to 
improve the signal to noise ratio of the measurements. The background subtracted 
outputs are combined two by two (with opposite signs) and normalized as follows, in 
order to remove scintillation noise which is common to both normal and 
complimentary outputs,

\begin{equation}
I_{ab} = \frac{I_1/\bar{I}_1 - I_2/\bar{I}_2} {2},
\end{equation}
and correspondingly for the other outputs. Here $\bar{I}$ denotes the mean over the entire scan. 

The power spectra of the resultant three outputs $I_{ab}$, $I_{bc}$ and $I_{ac}$ are computed 
and the fringe power for each of these three outputs is estimated by integrating the power (P)
under the fringe profile after background power subtraction \citep{Baldwin96}. The fringe power is proportional 
to the visibility-squared (V$^2$). The target V$^2_{target}$ is calibrated by measuring the fringe 
power for a nearby calibrator of known V$^2_{calib}$ under the same instrumental 
configuration and by taking the ratio. i.e.

\begin{equation}
V^2_{target} = V^2_{calib}  \left(\frac{P_{target}}{P_{calib}}\right)
\end{equation}

The closure phase is the sum of the fringe phases simultaneously observed 
on three baselines forming a closed triangle and is insensitive to phase errors induced by the turbulent atmosphere or optics 
\citep{Jennison58}. If the phase errors introduced into the three beams are 
$\delta_a$, $\delta_b$ and $\delta_c$, then the observed fringe phase between 
baselines a and b can be written as  
\begin{equation}
\phi_{ab} = \psi_{ab}+\delta_b-\delta_a
\end{equation}
Here $\psi_{ab}$ is the true object fringe phase between baselines a and b.

The observed closure phase $\Phi_{cl}$ is equal to the true object closure phase,
$\psi_{ab}+\psi_{bc}+\psi_{ca}$, to within the measurement noise, as follows:

\begin{equation}
\Phi_{cl} = \phi_{ab}+\phi_{bc}+\phi_{ca} +\rm{noise} 
\end{equation}
\begin{equation}
	= \psi_{ab}+\delta_b-\delta_a + \psi_{bc}+\delta_c-\delta_b +
		 \psi_{ca}+\delta_a-\delta_c +\rm{noise}
\end{equation}
\begin{equation}
        = \psi_{ab}+\psi_{bc}+\psi_{ca} +\rm{noise}
\end{equation}

We estimate closure phase as the phase of the bispectrum \citep{Weigelt77} of 
simultaneous fringes obtained with the three baselines. The instrumental 
closure phase is estimated using a nearby calibrator and subtracted from 
the raw closure phase of the target to give a calibrated target closure 
phase. 

Typical one-sigma formal errors in our uncalibrated closure phase and 
V$^2$ measurements are $\sim$ 0.2$^{\rm o}$ and $\sim$ 2\% respectively.
The formal errors are estimated from the scatters of the 500 fringe scans. In the case of 
V$^2$ errors, the error due to background power subtraction is also incorporated into 
the formal error. The calibration process adds up additional errors and the measurement error is estimated as the square-root of the sum of the squares of the formal and calibration errors. The one-sigma measurement error is reported in Tables~\ref{result} \& \ref{negResult}.
We estimate calibration errors by observing calibrators under same observing condition 
and calibrating one calibrator with the other after accounting for the finite sizes of both 
calibrators.
The estimated one-sigma calibration error is 5\% for the V$^2$ measurements 
and 0.5$^\circ$ for closure phase measurements. 
There could be unaccounted systematic errors in our measurements.
For the purpose of this article, we adopt a systematic error of 2$^\circ$ for our closure phase measurements. The total error is estimated as the square-root of the sum of the squares of the formal, calibration and systematic errors. If the measured closure phase is less than twice the total error in the measurement then we call it essentially a non-detection of asymmetry. However, if it is 
larger, we call it positive detection. 
Further discussion on our visibility and closure phase measurements could be found in \citet{Ragland04}.

\section{Results}

Targets with centro-symmetry should give
a closure phase of either zero or $\pm180^{\rm o}$ depending on how many 
baselines are beyond the 1st, 2nd, etc nulls. 
The majority of our targets show zero closure phase. 
However, 14 of the 56 have non-zero closure phase. 
Of these, 12 are Mira stars, and 4 are SR/Irr variable carbon stars. 
Among the 12 Mira stars, all but $\chi$ Cyg 
are oxygen-rich Mira stars; $\chi$ Cyg is classified as an S star. The 
frequency of 
asymmetry from our studies is 34\% in 
Mira stars, 17\% in SR variables, 33\% in Irr variables and thus 29\% in our 
entire sample of AGB stars. In terms of 
chemistry, the frequency of asymmetry is 33\% in carbon stars and 
27\% in oxygen-rich stars.

\begin{figure}[bthp]
\centering
\includegraphics[width=0.98\hsize]{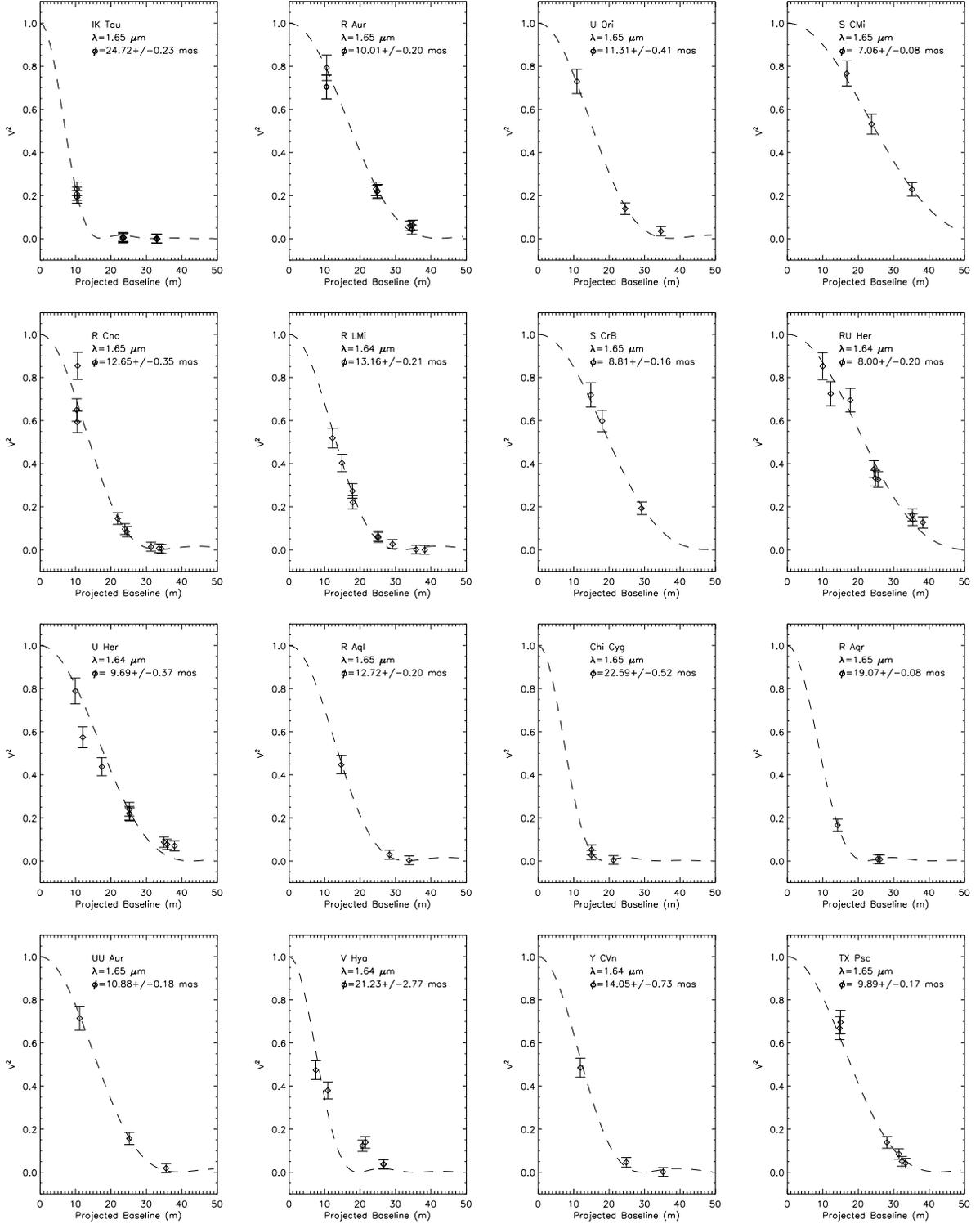}
\caption{{\footnotesize Visibility-data fitted with an UD model for the targets with positive asymmetry detection from our observations. UD models fits the data very well except for V Hya. In the case of V Hya, the derived size is possibly the size of the dust shell rather than that of the central star.
}}
\label{resultFig}
\end{figure}

Table~\ref{result} gives the measured closure phases along with observational and target information for these 14 targets that show measurable asymmetry 
from our observations. We have included in this table one observation each for R Cnc, R LMi \& V Hya in which no asymmetry is detected since we use these observations for size estimation. All targets (except R Aur, $\chi$ Cyg \& UU Aur) have follow up measurements taken with all three narrow band filters to characterize the observed asymmetry. These results will be presented elsewhere. The V$^2$ data for these targets are fitted with uniform disk (UD) 
models (figure~\ref{resultFig}) and the measured angular sizes are also given in Table~\ref{result}. Table~\ref{negResult} lists targets for which we have not detected asymmetry from our survey. The calibrators used for our observations are also listed in these tables.
The angular sizes for most of the calibrators are taken from \citet{Wesselink72}. Typically, we use calibrators with angular sizes less than 3 mas for our measurements. However, during the early part of the survey, we had to use larger calibrators because of low throughput of the instrument. We adopted the measured sizes from interferometric technique for  $\alpha$ Vul and $\gamma$ Sge \citep{Hutter89}, and from lunar 
occultation measurements for UU Aur \citep{Bohme78}. We estimated the angular size of 7 Peg from V and K magnitudes \citep{vanBelle99}.
%

\begin{deluxetable}{lclrllrlrrr}
\rotate
\tablewidth{0pt}
\tablecaption{Derived closure phase and UD diameters in the H band for targets with detected asymmetry.
\label{result}
}
\tablehead{
                           &                &
&\colhead{Calib.} &
\colhead{CGCVS}  	   & \colhead{AAVSO}&
&	   &  &\\

\colhead{Target}           & \colhead{Date}  &
\colhead{Calibrator}&
\colhead{diam.} &
\colhead{Period}	   & \colhead{Period}&
\colhead{Phase}&
\colhead{Sp. Type}  	   &
\colhead{B$_{max}$}& \colhead{$\theta_{UD}$} & \colhead{$\phi_{cp}$\tablenotemark{5}}\\

                &	UT	  & 
&
\colhead{(mas)}&
\colhead{(days)}  & \colhead{(days)}  & 
&&\colhead{(cm)}& \colhead{(mas)} & \colhead{(deg)}}
\startdata
\multicolumn{6}{l}{\underline{\bf (a) Mira Stars:}}\\
&&&&&&&&&&\\
IK Tau\tablenotemark{3}     &19Jan03& 63 Ari& 2.6& 470   &       &0.4& M6e-M10e         &3282.1&24.72$\pm$0.23&117.3$\pm$1.0\\
               &29Jan03&&&&&&&3282.3&&162.4$\pm$17.4\\
               &30Jan03&&&&&&&3309.1&&152.8$\pm$2.9\\  
               &       &&&&&&&3287.0&&144.4$\pm$3.2\\      
R Aur\tablenotemark{3}      & 19Jan03 & HD 31312& 2.6& 457.5 & 452.5 &0.4& M6.5e-M9.5e      &3461.5&10.01$\pm$0.20&-7.0$\pm$0.5\\ 
               &25Jan03&&&&&&&3416.9&&-6.6$\pm$0.6\\ 
               &28Jan03&&&&&&&3487.0&&-6.7$\pm$0.6\\              
U Ori\tablenotemark{3}      & 28Jan03    & 40 Ori  & 2.2& 368.3 & 372.3 &0.1& M6e-M9.5e           &3461.6&11.31$\pm$0.41&-8.6$\pm$0.6\\
S CMi          &08Mar03&27 Mon  & 2.4& 332.9&329.6&0.4&M7e    &3520.4&7.06$\pm$0.08&6.0$\pm$0.6\\  
R Cnc\tablenotemark{3}      & 23Jan03 & $\omega$ Hya & 2.5& 361.6 & 362.3 &0.3& M6e-M9e
&3354.7&12.65$\pm$0.35&6.7$\pm$0.6\\ 
               &      &&&&&&&3131.8&&-1.0$\pm$0.7\tablenotemark{1}\\ 
               & 28Jan03 &&&&&&&3424.7&&9.6$\pm$0.7\\ 
R LMi\tablenotemark{2,3} & 07Mar03 & $\omega$ Hya & 2.5& 372.2 & 375.4 &0.5& M7e    
&2916.8&13.16$\pm$0.21&2.6$\pm$0.5\tablenotemark{1}\\ 
               & 09Mar03 &&&&&&&3825.4&&-5.6$\pm$1.2\\ 
               & 12Mar03 &&&&&&&3582.7&&-31.5$\pm$0.8\\ 
S CrB          & 07Mar03    & HR 5464 & 2.7& 360.3 & 365     &0.5& M7e          
&2914.3& 8.81$\pm$0.16&  5.5$\pm$0.5\\ 
RU Her\tablenotemark{2}     & 10Mar03 & 51 Her  & 2.5& 484.8 & 494.1 &0.1& M6e-M9
&3823.6& 8.00$\pm$0.20&-7.2$\pm$0.7\\ 
               & 11Mar03 &&&&&&&3539.3&&-6.6$\pm$0.6\\ 
               & 13Mar03 &&&&&&&3530.0&&-11.0$\pm$0.7\\ 
U Her\tablenotemark{3}      & 28Jun02 &$\kappa$ Ser& 6.6& 406.1 & 406.5 &0.0& M6.5e-M9.5e
&3657.9&9.69$\pm$0.37&20.9$\pm$1.6\\ 
R Aql\tablenotemark{3}      & 27Jun02    & $\gamma$ Sge & 6.9& 284.2 & 272.9 &0.7& M5e-M9e          &3385.9&12.72$\pm$0.20&-26.6$\pm$0.8\\ 
$\chi$ Cyg\tablenotemark{3} & 29May02 &$\alpha$ Vul& 5.1& 408.1 & 404.8 &0.2& S6,2e-S10,4e     &2121.9&22.59$\pm$0.52&168.7$\pm$1.4\\ 
R Aqr\tablenotemark{3}      & 30Oct02    & $\iota$ Cet & 4.7& 387.0 & 386.1 &0.3& M5e-M8.5e+pec    &2614.3&19.07$\pm$0.08&-11.0$\pm$0.9\\ 
\multicolumn{6}{l}{\underline{\bf (b) Semi-regular Variables:}}\\
&&&&&&&&&&\\
UU Aur\tablenotemark{3}     & 30Jan03    & HD 61603& 2.9& 234   &(457.5)&   &  C5,3-C7,4 
&3554.4& 10.88$\pm$0.18&-15.5$\pm$0.6\\ 
V Hya\tablenotemark{2}      &10Mar03 &$\alpha$ Crt & 2.9& 530.7 &       &0.1& C6,3e-C7,5e 
&2655.2& 21.23$\pm$2.77\tablenotemark{4}&17.4$\pm$0.7\\ 
               & 11Mar03 &&&&&&&2666.6&&0.4$\pm$0.6\tablenotemark{1}\\ 
Y CVn\tablenotemark{3}      & 28Jan03    & HR 5464 & 2.7& 157   &       &   &C5,4J        
&3487.7&14.05$\pm$0.73&-26.9$\pm$1.8\\ 
\multicolumn{6}{l}{\underline{\bf (c) Irregular Variables:}}\\
&&&&&&&&&&\\
TX PSc\tablenotemark{3}&29Oct02&$\iota$ Cet &4.7&&(255.5)&&CII...&3336.5&9.89$\pm$0.17&-4.6$\pm$0.7\\ 
          &30Oct02&        &&&&&         &3234.3&&-1.5$\pm$0.7\tablenotemark{1}\\ 
\enddata

\tablenotetext{1} {This observation doesn't show asymmetry.}
\tablenotetext{2} {Observed with the narrow band filter at 1.64$\mu$m.}
\tablenotetext{3} {At least one of the baseline resolved this target to the level of below 5\% in V$^2$.}
\tablenotetext{4} {UD model failed to fit the data. The derived value probably gives the diameter of the dust shell rather than diameter of central star.}
\tablenotetext{5} {The uncertainties indicates random errors only (see text).}
\end{deluxetable}


\begin{figure}[bthp]
\centering
\includegraphics[width=0.6\hsize]{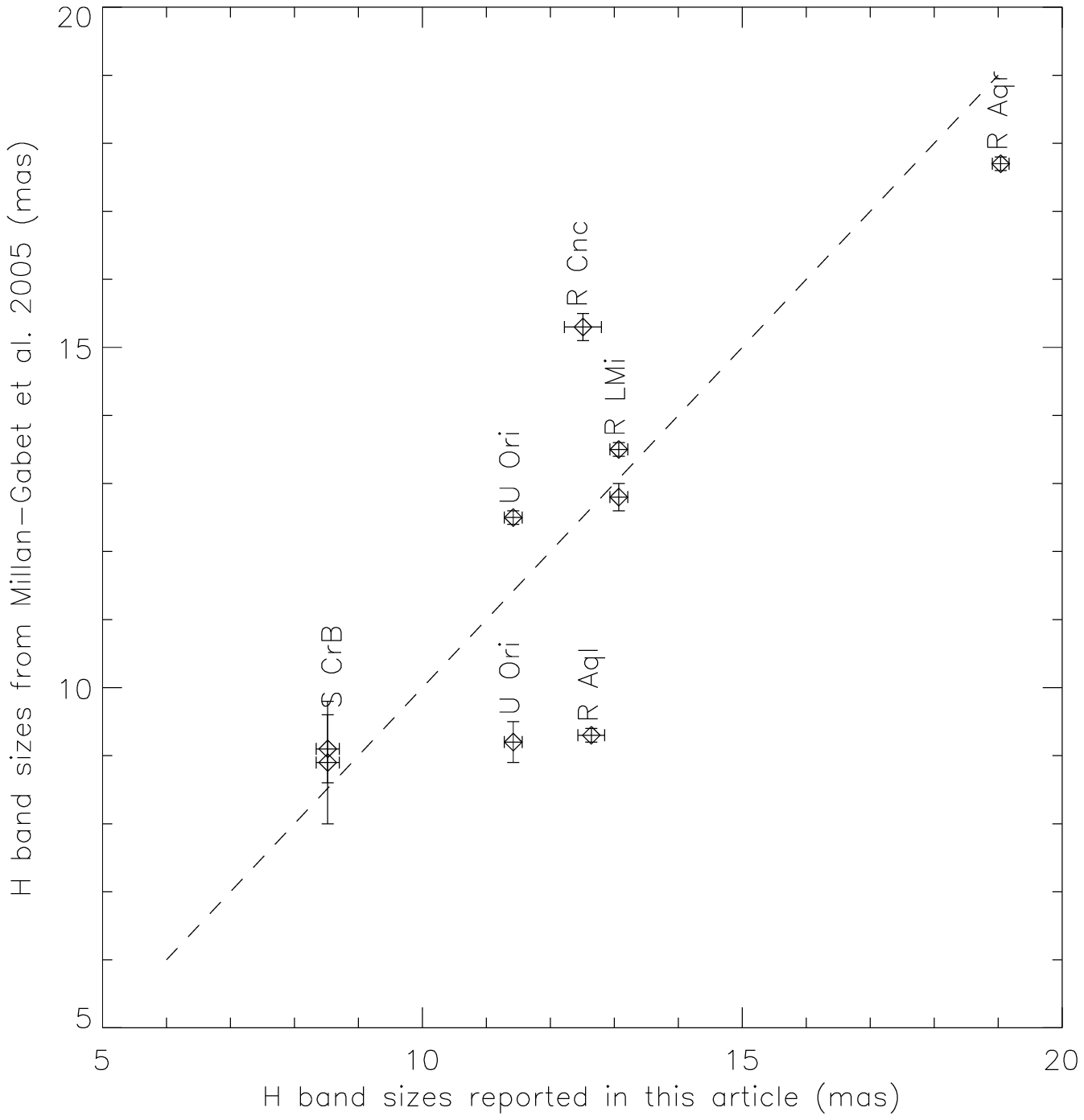}
\caption{{\footnotesize Shows the comparison of sizes reported in this article with those reported in the literature.
}}
\label{validate}
\end{figure}

Six of our targets with detected asymmetry, namely, U Ori, R Cnc, R LMi, S CrB, R Aql, R Aqr have earlier H band size measurements \citep{Millan05} taken within $\pm$ 0.2 pulsation phase with respect to the pulsation phase of our measurements for these stars. We have plotted our size measurements against the measurements by these authors for these six targets in Figure~\ref{validate}. 
The scatter in this figure 
is comparable to the scatter among measurements of a given star at 
multiple epochs in a single program, and thus probably either 
signifies actual size variation at the source or that the 
interpretation of the observations (UD) is too simple.
Several targets have earlier K band angular size measurements (\citet{vanBelle96}; \citet{vanBelle02}; \citet{Millan05}; \citet{Mennesson02}; \citet{Dyck96}).

\begin{deluxetable}{lclrrccrrr}
\rotate
\tablewidth{0pt}
\tablecaption{Closure phase measurements for targets with no detectable asymmetries
\label{negResult}
}
\tablehead{
                           &                &&
\colhead{CGCVS}  	   &\colhead{AAVSO}&
	   &&  &\\

\colhead{Target}           & \colhead{Date}  &
\colhead{Calib.}&
\colhead{Period}	   &\colhead{Period}&
\colhead{Phase}&
\colhead{Sp. Ty.}  	   &\colhead{B$_{max}$}& \colhead{$\theta_{UD}$} & \colhead{$\phi_{cp}$\tablenotemark{15}}\\

                &	UT	  & 
&
\colhead{(days)}  & 
		  &&\colhead{(cm)}& \colhead{(mas)} & \colhead{(deg)}}
\startdata
\multicolumn{6}{l}{\underline{\bf (a) Mira Stars:}}\\
&&&&&&&&&\\
U Per  &30Oct02&51 And &320.3&318.7&0.5&M6e    &3607.2&5.84\tablenotemark{2}&1.4$\pm$1.1\\ 
R Tri  &29Jan03&24 Per &266.9&264.8&0.9&M4IIIe &3548.7&4.46\tablenotemark{3}&2.8$\pm$0.5\\ 
RT Eri &31Jan03&HR 1543&370.8&376.5&0.9&M:e    &2527.9&6.3\tablenotemark{4}&1.0$\pm$0.6\\ 
R Lep\tablenotemark{1}  &31Jan03&HR 1543&427.1&437.8&0.8&CIIe...&2904.8&11.50\tablenotemark{5}&-3.9$\pm$0.6\\ 
RU Aur &29Jan03    &51 Ori &466.5&464.3&0.7&M8     &3553.5&3.7\tablenotemark{4}&-0.9$\pm$0.6\\ 
X Aur  &29Oct02&$\upsilon$ Aur&163.8&166.1&0.2&K2     &3664.0&1.8\tablenotemark{4}&2.5$\pm$1.1\\ 
	     &23Jan03&HD 31312&    &&0.2&       &3378.4&       &-0.2$\pm$0.6\\ 
       &25Jan03&       &     &&0.2&       &3429.5&       &-0.9$\pm$0.6\\ 
       &28Jan03&       &     &&0.2&       &3507.5&       &-2.1$\pm$0.6\\ 
V Mon  &29Jan03&       &340.5&332.7&   &M6e    &2771.1&5.6\tablenotemark{4}&2.6$\pm$0.6\\ 
W Cnc  &28Jan03&$\omega$ Hya&393.2&391.8&0.3&M7e    &3415.2&4.7\tablenotemark{4}&-0.7$\pm$0.6\\ 
X Hya  &30Jan03&28 Hya &301.1&301.2&0.6&M7e    &2451.7&5.0\tablenotemark{4}&-0.2$\pm$0.6\\ 
R LMi  &28Jan03&HR 5464&372.2&375.4&0.3&M7e    &3530.8&13.2\tablenotemark{2}&-2.6$\pm$0.7\\ 
V Boo  &29May02&$\rho$ Boo&258.0&261.1&0.6&M6e    &2100.6& 5.2\tablenotemark{4}&-0.7$\pm$0.7\\ 
	     &28Jan03&HR 5464&     &     &0.6&       &3448.6&        &0.2$\pm$0.6\\ 
S CrB  &29May02&52 Boo &360.3&365  &0.7&M7e    &2121.3& 9.1\tablenotemark{2}&0.3$\pm$0.8\\ 
       &28Jan03&HR 5464&     &     &0.4&       &3509.1& 8.9\tablenotemark{2}&1.5$\pm$0.6\\ 
S Ser  &28Jun02&$\kappa$ Ser&371.8&373.7&0.9&M5e    &3557.1& 5.35\tablenotemark{6}&-1.3$\pm$1.5\\ 
BG Ser &29Jun02&$\kappa$ Ser&143  &386.1&0.7&M6me...&3042.7& 6.71\tablenotemark{6}&0.1$\pm$0.9\\ 
			 &30Jan03&110 Vir&     &     &0.3&       &3328.4&         &0.7$\pm$0.7\\ 
			 &       &       &     &     &   &       &3305.8&         &-0.2$\pm$0.7\\ 			 
R Ser  &27May02&$\kappa$ Ser&356.4&355.6&0.4&M7IIIe &2089.8& 7.6\tablenotemark{2}  &0.0$\pm$1.0\\ 
       &28May02&$\kappa$ Ser&     &     &0.4&       &2052.9&         &0.4$\pm$0.8\\ 
V CrB  &30Jan03&HR 5464&357.6&361.3&0.9&N...   &3271.8& 7.26\tablenotemark{5} &-1.5$\pm$0.6\\ 
RU Her &25May02&HR 5947&484.8&494.1&0.5&M7e... &2116.5& 8.71\tablenotemark{6}&1.3$\pm$0.9\\ 
RT Oph &28Jun02&$\beta$ Oph&426.3&424.8&0.0&M7 &3500.1& 6.52\tablenotemark{6}&-1.0$\pm$1.0\\ 
X Oph  &26May02&$\gamma$ Aql&328.9&341.1&0.4&K1III+...&1968.1&12.97\tablenotemark{3}&-0.6$\pm$0.8\\ 
R Aql  &26May02&$\gamma$ Aql&284.2&272.9&0.5&M7IIIe&1944.2&9.3\tablenotemark{2}&-1.1$\pm$0.8\\ 
W Aql  &28May02&$\gamma$ Aql&490.4&480.8&0.1&S:...&1650.0&11.08\tablenotemark{5}&1.2$\pm$1.0\\ 
RT Aql &24Jun02&$\gamma$ Aql&327.1&325.7&0.0&M7e  &3488.8& 7.24\tablenotemark{6}&1.9$\pm$0.8\\ 
BG Cyg &29Jun02&$\gamma$ Sge&228  &285.4&0.1&M7e  &3793.5& 4.14\tablenotemark{6}&-1.8$\pm$1.0\\ 
RR Aql &28May02&$\gamma$ Aql&394.8&398.7&1.0&M7e  &1795.9&10.73\tablenotemark{6}&0.2$\pm$1.0\\ 
U Cyg  &28Jun02&$\beta$ Oph &463.2&469  &0.0&R... &3732.1& 7.05\tablenotemark{5}&-3.5$\pm$1.4\\ 
V1426 Cyg&30Oct02&$\rho$ Cyg&470  &481.2&0.9&C    &3816.0&10.8\tablenotemark{7} &-0.8$\pm$0.9\\ 
R Peg  &31Oct02&7 Psc  &378.1&378.2&0.8&M7e       &3579.7& 7.0\tablenotemark{2} &-1.9$\pm$1.2\\ 
\multicolumn{6}{l}{\underline{\bf (b) Semi-regular Variables:}}\\
&&&&&&&&&\\
$\rho$ Per&21Nov02&24 Per       & 50&&&M4II      &&15.53\tablenotemark{8}&1.3$\pm$3.5\\ 
W Ori  &31Oct02&40 Ori&212&(446)&&CII...    &3394.3& 9.7\tablenotemark{7}&5.2$\pm$1.8\\ 
CE Tau &30Oct02&HD 33554&165&&&M2Iab:  &3717.4& 9.1\tablenotemark{9}&-4.1$\pm$1.2\\ 
X Cnc  &19Jan03&$\omega$ Hya&195&379.6&0.3&CII...&3527.5& 7.62 \tablenotemark{10}&1.8$\pm$0.6\\ 
       &25Jan03&       &   &&0.3&      &3473.1&         &-0.4$\pm$0.6\\ 
       &28Jan03&       &   &&0.3&      &3410.5&         &2.1$\pm$0.6\\ 
T Cnc  &23Jan03&$\omega$ Hya&428&499.5&   &N...  &3447.1& 6.6\tablenotemark{11}&-2.0$\pm$0.6\\ 
       &28Jan03&       &   &     &   &      &3409.6&         &0.3$\pm$0.6\\ 
U Hya  &25Jan03&$\alpha$ Crt&450&&&CII...&2902.0&10.8\tablenotemark{4}&-2.4$\pm$0.7\\
			 &30Jan03&HR 5464     &   &&&      &2489.6&        &-2.8$\pm$0.6\\
V Hya  &23Jan03&$\alpha$ Crt&530.7&&1.0&C...  &2640.1&13.0\tablenotemark{4}&2.1$\pm$0.6\\ 
Y CVn  &29May02&HR 4690&157&&&CIab:... &2022.7&11.6\tablenotemark{7}    &-3.6$\pm$1.0\\ 
RT Vir &28May02&$\upsilon$ Boo&155&&&M8III    &1902.3&12.38\tablenotemark{12}&0.7$\pm$0.7\\ 
SW Vir &28May02&$\upsilon$ Boo&150&155.4&&M7III    &1818.9&16.24\tablenotemark{12}&2.8$\pm$0.9\\ 
RX Boo\tablenotemark{1} &23May02&$\rho$ Boo&340&&&M7.5 &2112.2&17.48\tablenotemark{12}&-3.9$\pm$1.0\\ 
ST Her &23May02&HR 5763&148&&&M6s      &1945.5& 9.3\tablenotemark{13}&-0.1$\pm$1.0\\ 
X Her  &25May02&       &95 &&&M        &&12.1\tablenotemark{13}&\\ 
g Her\tablenotemark{1}&29May02&52 Boo  &89.2&&&M6III &2063.0&12.67\tablenotemark{12}&-2.5$\pm$1.0\\ 
R Lyr\tablenotemark{1} &25May02&HR 6695&46&&&M5III  &2056.4&13.3\tablenotemark{13}&0.3$\pm$0.9\\ 
V Aql  &28Jun02&$\beta$ Oph&353&377.1&&CII...&2870.1&10.1\tablenotemark{7}&-0.8$\pm$1.0\\ 
EU Del &26May02&$\gamma$ Sge&59.7&&0.7&M6III&2063.7& 9.8\tablenotemark{9}&-0.8$\pm$0.8\\ 
\multicolumn{6}{l}{\underline{\bf (c) Irregular Variables:}}\\
&&&&&&&&&\\ 
Del2 Lyr&25May02&HR 6695&&&&M4II&2099.3&10.32\tablenotemark{14}&-0.6$\pm$0.9\\ 
EPS Peg&25Jun02&7 Peg   &&&&K2Ib&3657.3& 7.7\tablenotemark{13}&-0.1$\pm$1.1\\ 
TX PSc\tablenotemark{1}&25Jun02&7 Peg&&(255.5)&& CII...&3571.0&11.2\tablenotemark{7}&4.9$\pm$1.6\\ 
\enddata
\tablenotetext{1}{At least one of the baseline resolve this target to the level of below 5\% in V$^2$.}\\
\tablenotetext{2}{\citet{Millan05}};\tablenotetext{3}{\citet{Thompson02}};\tablenotetext{4}{\citet{vanBelle99}};\tablenotetext{5}{\citet{vanBelle97}}; 
\tablenotetext{6}{\citet{vanBelle02}};\tablenotetext{7}{\citet{Dyck96}};\tablenotetext{8}{\citet{DiBen93}};\tablenotetext{9}{\citet{Dyck98}}; 
\tablenotetext{10}{\citet{Richichi93}};\tablenotetext{11}{\citet{Richichi91}};\tablenotetext{12}{\citet{Mennesson02}};\tablenotetext{13}{\citet{Dyck96B}};\tablenotetext{14}{\citet{Sudol02}}\\
\tablenotetext{15}{The uncertainties indicates random errors only (see text).}

\end{deluxetable}

Out of 56 AGB stars, 16 are well resolved (i.e. at least one baseline 
gives a V$^2$ measurement less than 5\% of the point source value) 
from our observations. Among these 16 targets, 12 of them show asymmetry from 
our observations. Thus, if we consider only well resolved targets, 75\% of AGB stars
show asymmetry. Targets from our measurements that are well resolved are marked in Tables~\ref{result} \& \ref{negResult} (see the footnote). Interestingly, all well resolved targets that don't show asymmetry (except R Lep), namely, g Her, R Lyr \& RX Boo are SR variables; R Lep is a carbon Mira. Thus, if we consider only well-resolved oxygen rich Mira stars then our asymmetry detection is 100\% and in the case of SR variables the success rate is 40\%.
The well-resolved Mira variable R LMi did not show asymmetry from our January 2003 observations. However, we detected asymmetry in this target during March 2003 observations. Similarly, the well-resolved Irr variable TX Psc did not show asymmetry from our June 2002 measurements. However, we detected asymmetry in TX Psc from our follow-up observations taken in October 2002.

At least five targets with detected asymmetry, namely, S CrB, RU Her, R Aql, V Hya and Y CVn  have earlier measurements taken with relatively shorter baselines 
and we did not detect asymmetry from these measurements. 

In order to understand the role of angular resolution on the asymmetry detection, we derived the number of pixel elements (N$_{pix}$) in an imaging sense, defined as the angular diameter divided by the angular resolution ($\lambda$/2$B_{max}$) for all targets.
Here B$_{max}$ is the maximum baseline of our observations listed in Tables~\ref{result} \& \ref{negResult} for all our targets. We plotted our measured closure phase values against number of pixel elements in Figure~(\ref{cPhase}). This figure clearly shows that the positive closure phase cases are those that have pixel elements close to or greater than unity, meaning they are well resolved. This suggests that the detected asymmetry features are probably on the surface of the stellar disk 
or visible only in projection against the stellar disk (such as might be the case for patchy dust opacity at 1.5 to 2.5 stellar radii). 

\section{Simple Models}

We have assumed a 2-component brightness distribution model in order to find the simplest possible implications of the measured closure phase signal. This model consists of a uniform-disk star with an intensity distribution $\tilde{I}_p$($\vec{r}$) and an unresolved secondary component (bright spot, companion or dust clump) with an intensity distribution $\tilde{I}_s$($\vec{r}-\vec{\delta r}$), where 
$\vec{\delta r}$ is the separation vector between the optical centers of the components. The total intensity is $I$ = $\tilde{I}_p$($\vec{r}$) + $\tilde{I}_s$($\vec{r}-\vec{\delta r}$). The complex visibility is the Fourier transform of the brightness distribution. Thus, the complex visibility of this composite object could be written using the shift theorem for Fourier transforms as 
\begin{equation}
\hat{V}(\vec{g}) = \hat{V}^p({\vec{g}}) + \hat{V}^s({\vec{g}}) \hspace{0.1in} e^{i k\vec{g} . \vec{\delta r}}
\end{equation}
where k = 2$\pi$/$\lambda$, $\lambda$ is the wavelength of observation and $\vec{g}$ is the baseline vector $\vec{B}$.

The complex visibility for the baseline $\vec{B}_{AB}$ could be written as 
\begin{equation}
\hat{V}_{AB} = V^p_{AB}e^{i\phi^p_{AB}} + V^s_{AB} e^{i\phi^s_{AB}} e^{i k\vec{B}_{AB} . \vec{\delta r}}
\end{equation}

The visibility phase for the baseline B$_{AB}$ is
\begin{equation}
\phi_{AB} = \arctan \left[\frac{Im \hspace{0.05in}\hat{V}_{AB}}{Re \hspace{0.05in}\hat{V}_{AB}}\right]\\
\end{equation}
\begin{equation}
= \arctan \left[\frac{V^p_{AB}~\sin(\phi^p_{AB}) + V^s_{AB} ~\sin(\phi^s_{AB}+k\vec{B}_{AB} . \vec{\delta r})}{V^p_{AB}~\cos(\phi^p_{AB}) + V^s_{AB} ~\cos(\phi^s_{AB}+k\vec{B}_{AB} . \vec{\delta r})}\right]
\end{equation}
\begin{equation}
= \arctan \left[\frac{V^p_{AB}~\sin(\phi^p_{AB}) + V^s_{AB} ~{\sin(\phi^s_{AB}) \cos(k\vec{B}_{AB} . \vec{\delta r}) + V^s_{AB} \cos(\phi^s_{AB}) \sin(k\vec{B}_{AB} . \vec{\delta r})}}{V^p_{AB}~\cos(\phi^p_{AB}) + V^s_{AB} ~{\cos(\phi^s_{AB}) \cos(k\vec{B}_{AB} . \vec{\delta r}) - V^s_{AB} \sin(\phi^s_{AB}) \sin(k\vec{B}_{AB} . \vec{\delta r})}}\right]
\end{equation}

The individual components of the brightness distributions are assumed to be circularly symmetric. Hence, ~$\sin$($\phi^p_{AB}$) = ~$\sin$($\phi^s_{AB}$) = 0, but ~$\cos$($\phi^p_{AB}$) and ~$\cos$($\phi^s_{AB}$) can be +1 or -1, depending on details of the case. Thus,

\begin{equation}
\phi_{AB} = \arctan \left[\frac{V^s_{AB} ~\cos(\phi^s_{AB}) \sin(k\vec{B}_{AB} . \vec{\delta r})}{V^p_{AB}~\cos(\phi^p_{AB}) + V^s_{AB} ~\cos(\phi^s_{AB}) \cos(k\vec{B}_{AB} . \vec{\delta r}) }\right]
\end{equation}

\noindent {\bf Unresolved Secondary component:}

For an unresolved (i.e. a point source) $\phi^s_{AB}$ = 0. Thus,
\begin{equation}
\phi_{AB} = \arctan \left[\frac{V^s_{AB}  ~\sin(k \vec{B}_{AB} . \vec{\delta r})}{V^p_{AB} ~\cos(\phi^p_{AB}) + V^s_{AB}  ~\cos(k \vec{B}_{AB} . \vec{\delta r})}\right]
\end{equation}

For a uniform-disk star, 
\begin{equation}
\hat{V}^p_{AB} = \left[\frac{2 J_1(\pi N^{AB}_{pix}/2)}{\pi N^{AB}_{pix}/2}\right] I_p, 
\end{equation}
and 
\begin{equation}
V^p_{AB} ~\cos(\phi^p_{AB}) = \left[\frac{2 J_1(\pi N^{AB}_{pix}/2)}{\pi N^{AB}_{pix}/2}\right] I_p, 
\end{equation}

For an unresolved secondary component, 
\begin{equation}
V^s_{AB} = I_s,
\end{equation}
where $I_p$ and $I_s$ are the normalized star and secondary component intensities (i.e. $I_p$ + $I_s$ = 1), and

\begin{equation}
N^{AB}_{pix} = \frac{\theta_{UD}}{(\lambda/2B_{AB})}
\end{equation}
 
Combining eqn (13), (15) and (16), we get
\begin{equation}
\phi_{AB}  = \arctan \left[\frac{I_s ~\sin(k \vec{B}_{AB} . \vec{\delta r})}{I_p \left(\frac{2 J_1(\pi N^{AB}_{pix}/2)}{\pi N^{AB}_{pix}/2}\right) + I_s ~\cos(k \vec{B}_{AB} . \vec{\delta r})}\right]
\end{equation}

\noindent {\bf Resolved Secondary component:}

For a resolved Gaussian secondary component (such as dust clump) $\phi^s_{AB}$ = 0, and  
\begin{equation}
V^s_{AB} = \left[exp(-(\pi \beta N^{AB}_{pix}/ 2 \alpha)^2)\right] I_s 
\end{equation}
where, $\alpha$ = 2 $\sqrt{ln(2)}$ and $\beta$ is the ratio of the size of the primary component to the secondary. 

Thus,
\begin{equation}
\phi_{AB}  = \arctan \left[\frac{I_s ~exp(-(\pi \beta N^{AB}_{pix}/ 2 \alpha)^2)~\sin(k \vec{B}_{AB} . \vec{\delta r})}{I_p ~\left(\frac{2 J_1(\pi N^{AB}_{pix}/2)}{\pi N^{AB}_{pix}/2}\right) + I_s ~exp(-(\pi \beta N^{AB}_{pix}/ 2 \alpha)^2)~\cos(k \vec{B}_{AB} . \vec{\delta r})}\right]
\end{equation}


For a resolved uniform-disk secondary component (such as stellar companion) $\phi^s_{AB}$ = 0 or $\pm \pi$, and
\begin{equation}
\hat{V}^s_{AB} = \left[\frac{2 J_1(\pi \beta N^{AB}_{pix}/2)}{\pi \beta N^{AB}_{pix}/2}\right] I_s, 
\end{equation}
and 
\begin{equation}
V^s_{AB} ~\cos(\phi^s_{AB}) = \left[\frac{2 J_1(\pi \beta N^{AB}_{pix}/2)}{\pi \beta N^{AB}_{pix}/2}\right] I_s, 
\end{equation}

Now,
\begin{equation}
\phi_{AB}  = \arctan \left[\frac{I_s \left(\frac{2 J_1(\pi \beta N^{AB}_{pix}/2)}{\pi \beta N^{AB}_{pix}/2}\right) ~\sin(k \vec{B}_{AB} . \vec{\delta r})}{I_p \left(\frac{2 J_1(\pi N^{AB}_{pix}/2)}{\pi N^{AB}_{pix}/2}\right) + I_s \left(\frac{2 J_1(\pi \beta N^{AB}_{pix}/2}{\pi \beta N^{AB}_{pix}/2}\right)~\cos(k \vec{B}_{AB} . \vec{\delta r})}\right]
\end{equation}

\begin{figure}[bthp]
\centering
\includegraphics[width=0.95\hsize]{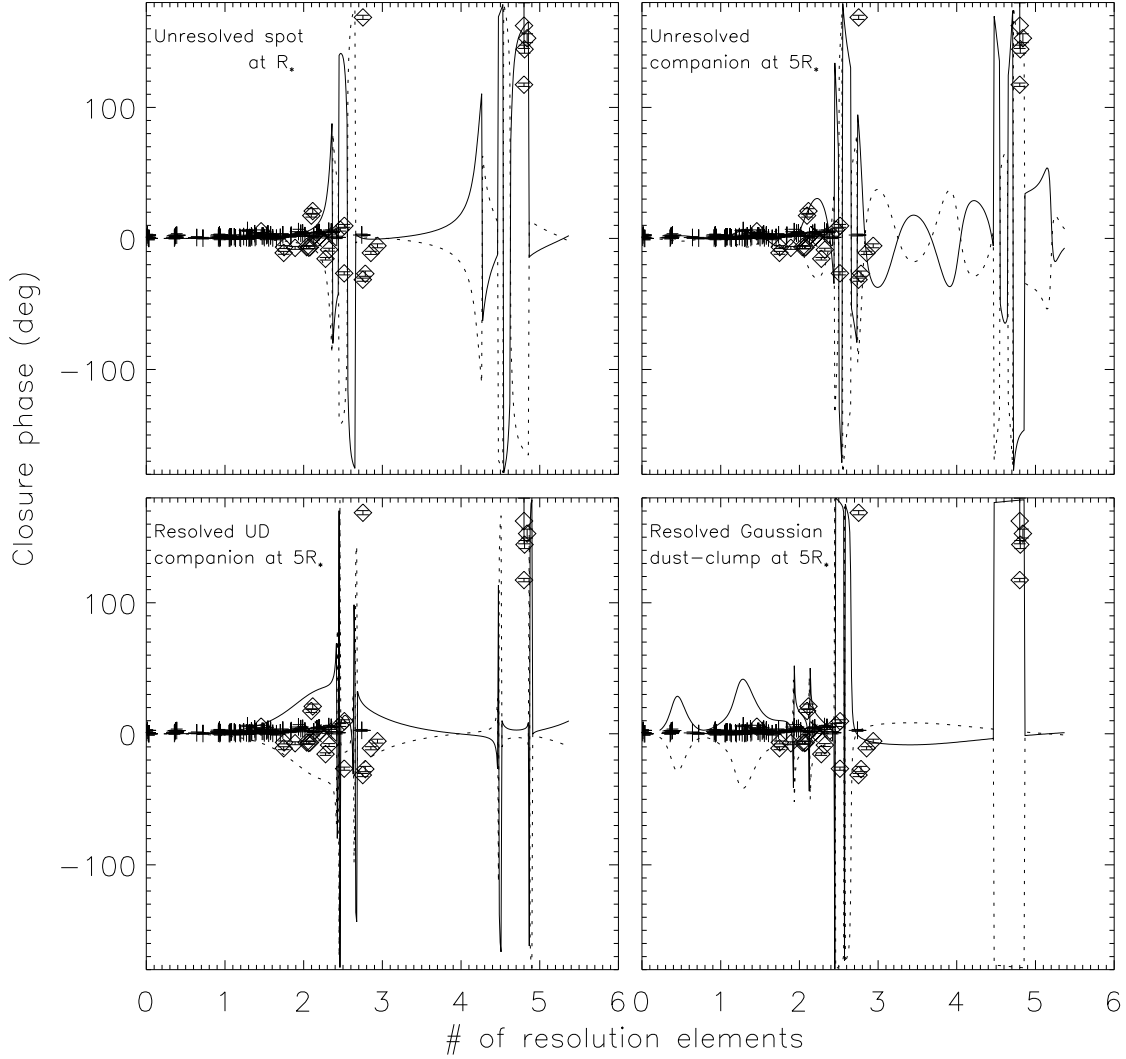}
\caption{{\footnotesize Measured closure phases are plotted against number of pixel elements (see the text). Targets with positive asymmetry detection are shown as diamond symbols and non-detection measurements are shown as plus symbols. Targets that are well resolved frequently show large closure phase. The solid lines refer models where the secondary feature is assumed at the position angle of 21.8$^{\rm o}$ (along the direction of the largest baseline of the IOTA array) and the dotted lines refer to models where the secondary feature is assumed at the position angle of 201.8$^{\rm o}$ (21.8$^{\rm o}$ + 180$^{\rm o}$). The flux of the unresolved secondary components are assumed to be 3\% and that of the resolved secondary component is assumed to be 30\%.
Top left: Unresolved spot at the edge of the stellar disc ($\delta r$ = $\theta_{UD}$/2);
Top right: Unresolved companion at 5 stellar radii ($\delta r$ = 2.5 $\theta_{UD}$); 
Bottom left: Resolved UD companion at 5 stellar radii ($\delta r$ = 2.5 $\theta_{UD}$); The diameter of the secondary is assumed to be 99\% of the diameter of the primary;
Bottom right: Resolved Gaussian dust clump at 5 stellar radii ($\delta r$ = 2.5 $\theta_{UD}$); The equivalent size of the secondary is assumed to be same as that of the primary.
}}
\label{cPhase}
\end{figure}

Using eqn (4), (18), (20) and (23), we generate closure phase models for the IOTA configuration A35B15C00 (i.e. B$_{CA}$ = 35 m; B$_{BC}$ = 15 m) which are shown with our data in Figure~\ref{cPhase}. We explored the parametric space with four cases, namely (1) surface-unresolved spot, (2) surface-unresolved companion, (3) surface-resolved companion and (4) surface-resolved Gaussian dust clump. The values assumed for various parameters of these models are primarily for the purpose of illustration. More thorough treatment of the physical parameters chosen in the models will be presented in the forthcoming article. We have shown two models for each of the four cases (totally eight models) in Figure~\ref{cPhase}. The two models differ only in the position angle of the secondary feature - one assumes the secondary feature at the position angle of 21.8$^{\rm o}$ and the other at the position angle of 201.8$^{\rm o}$ (i.e. a 180$^{\rm o}$ rotation). The reason for choosing this axis (along the direction of the largest baseline of the IOTA array) is that the closure-phase signal is approximately maximum when the secondary feature is assumed along this axis. The brightness distribution of the primary component is assumed to be an uniformly illuminated disk. The flux of the secondary component is assumed to be 3\% of the total flux for surface-unresolved secondary cases and 30\% for surface-resolved secondary component cases; the reason for choosing these flux values for the companion is that the corresponding models compare well with our closure phase measurements. The spot models (case 1) assume a surface-unresolved bright spot at the edge of the stellar disc ($\delta r$ = $\theta_{UD}$/2). The surface-unresolved binary models (case 2) assume a surface-unresolved companion or dust clump at 5 stellar radii ($\delta r$ = 2.5 $\theta_{UD}$). The surface-resolved binary models (case 3) assume companion with $\beta$ = 0.99 (UD angular size of the secondary component is 99\% of the primary) at 5 stellar radii ($\delta r$ = 2.5 $\theta_{UD}$). 
If companion stars provide enough light to produce detectable 
asymmetry near maximum light, then they should also produce a wider, 
flattened, minima in the light curves \citep{Merrill56}.  While at least 
two of the stars are known binaries and relatively mild symbiotic 
systems (R Aqr and o Ceti) none of the stars show a filled-in minimum 
on the AAVSO light curves.
The dust models (case 4) assume a Gaussian shaped dust clump ($\beta$ = 1; same equivalent size as the primary component) at 5 stellar radii ($\delta r$ = 2.5 $\theta_{UD}$). As can be seen in Figure~\ref{cPhase}, the unresolved spot models compare well with the observed data and the unresolved companion models are not as good as the unresolved spot models in explaining the observed closure phase data.
The resolved secondary feature models (case 3 \& 4) show that the secondary components have to be significantly brighter (may be physically unreasonable) in order to produce detectable closure phase signals and even so, the fits at low N$_{pix}$ are poor.  

\section{Discussion}

The significance of the present results 
(i.e., that 1/3 are asymmetric and 2/3 are not)
depends on what causes the asymmetry, and 
that is not yet known.  As noted in the 
introduction, although planetary nebulae are 
predominantly bipolar, a large fraction 
(possibly even all) of proto-PN are axisymmetric 
inside an apparently spherical AGB wind remnant, 
suggesting that the PN asymmetry has arisen only 
after or as the star left the AGB.  Asymmetries 
have been reported before for Miras, but these 
are mostly for isolated examples, or for maser 
emission that is very sensitive to the local 
conditions and thus will tend to exaggerate any 
physical departure from spherically symmetric 
flow. The present results, referring to intensity 
near the stellar flux maximum, are much less 
sensitive to small variations in the conditions.
It's worth mentioning here that the non-detections don't 
preclude asymmetries. It could be that they are just 
not resolved or the asymmetries are too small.

The SiO maser emission also arises well above the 
photosphere (e.g. Humphreys et al 2002; \citet{Cotton04}); in fact, 
there is, to our knowledge, no report yet of 
asymmetry that can be assigned unambiguously to 
the stellar surface.  With multiple narrow-band 
measurements at carefully selected wavelengths, 
Perrin et al (2004) have shown that it is 
possible to disentangle photospheric and shell 
contributions.  The results reported here suggest 
that we will also be able to sort out some shape 
information in the next generation of 
observations as well as separate the photospheric 
contribution from the circumstellar one.  

There is a considerable literature concerning the non-circular and 
non-spherical symmetry common among planetary nebulae; see, for 
example, the proceedings of Asymmetric Planetary Nebulae III (2004, 
ASP Conf.~Ser.~313).  Mechanisms may be roughly divided into deep and 
superficial.  Deep mechanisms include internal convective structure 
with large convection cells (proposed by Schwarzschild (1975) on the 
basis of simple scaling arguments), non-radial pulsation, and/or 
rotation.  

These stars have massive envelopes and large radii; no reasonable reservoir 
of angular momentum other than incorporation of a relatively massive 
($>$~0.1 M$_\odot$) companion will provide sufficient angular momentum for 
rotational asymmetry or the usual non-radial-pulsation associated with rotation.  
Large convective cells might stimulate non-radial modes in the absence of 
significant rotation, or might lead to modulations in the surface 
brightness from rising or falling elements.  Evolutionary models show 
a radiative layer above the convective layer in these stars \citep{Ostlie86}
and the scale heights at the photosphere are only a percent or 
so of R$_*$; these facts suggest that convective modulation will be on a 
smaller scale and with less contrast than is needed to explain these 
observations, but more detailed modeling should be done before the 
possibility of convection-based modulation is ruled out.

Superficial or 
atmospheric mechanisms include magnetic structures (e.g. Soker \&
Zoabi 2002; Blackman et al. 2001), discrete dust cloud formation as 
for R CrB stars, and interaction of a planet or companion with the 
stellar wind (Struck, Cohanim \& Willson 2004; Mastrodemos \& Morris 1998, 1999). 
The conclusion that perhaps all of the Miras show 
some asymmetry while only about half of the 
non-Miras do may be understood in a couple of 
ways.  Miras comprise a well-defined subset of 
long period variables, namely those with large 
visual amplitudes, relatively regular variation, 
cool effective temperatures, and moderate 
progenitor masses.  SR classes differ in visual 
amplitude (SRa), degree of irregular variation 
(SRb), warmer effective temperature (SRd), and higher progenitor 
mass (SRc).  Within each of these classes there 
are further probable subclasses.  Our current 
understanding is that the high visual amplitude 
is partly the result of variable atmospheric 
opacity \citep{Reid02}; this variable opacity is closely tied 
to the fact that these stars are losing mass at a 
high rate ($>$10$^{-7}$ M$_\odot$ per year up to ~10$^{-5}$ 
M$_\odot$ per year) so the evolutionary status of 
Miras is that they are stars entering the final 
"superwind" (massive outflow) stage on the AGB 
(e.g. review Willson 2000).  A number of stars 
initially classified as Miras are reclassified as 
SRb when their light curves develop 
irregularities, making the boundary between these 
two classes somewhat fuzzy.  Similarly, most 
relatively regular carbon star LPVs have smaller 
amplitudes in the visual than do the oxygen-rich 
stars, and this may be telling us more about the 
sources of atmospheric opacity than about the 
evolutionary state, so that the SRa-Mira boundary 
is also fuzzy.  Thus one interpretation would be 
that the asymmetry shows up when there is a 
sufficiently massive outflow to produce the large 
Mira amplitude for the oxygen-rich stars, and 
that the SR variables with asymmetries are those 
with different visual opacity but similarly 
massive outflows.  Either the same mechanism 
leads to outflow and asymmetry (e.g. non-radial 
pulsation) or the outflow sets up conditions for 
asymmetry to be seen.  In the first case, the 
non-radial structure originates at the 
photosphere; in the second, with aperiodicities 
in the outflow.

Most Miras are surrounded by translucent 
``molecular shells'', a locus in the outflow where 
molecules and probably dust provide high local 
opacity, whose IR and visual optical depth is on 
the order of 1 (0.1 to several) - Perrin et al (2004).
The physics of dust formation in the context of 
large-amplitude pulsation and consequent shocks 
is reviewed in Willson (2000).
Dust grains nucleated in the refrigerated zone 
between shocks may require several pulsation 
cycles to grow to sufficient size to generate an 
outflow, and this would naturally lead to 
critical dust levels appearing in different 
cycles at different positions around the star. 
Whether by this or another mechanism, the 
translucent shell is likely to have a patchy 
opacity, allowing more of the photospheric light 
through in some places than in others.  Thus a 
plausible explanation for the possibly universal 
asymmetry in Miras would be the formation of an 
inhomogeneous translucent molecular screen around 
1.5 to 2.5 stellar radii. 

In conclusion, we carried out a survey of AGB stars with the IOTA 3-telescope imaging interferometer at near-infrared
wavelengths, searching for asymmetry in their flux distributions. We find 
that 29\% of our sample show asymmetry. If we restrict the sample to only well 
resolved targets, then 75\% of AGB stars, 100\% of oxygen-rich Mira stars show 
asymmetry from our observations. 
On this basis, we hypothesize that all Mira stars might show detectable asymmetry if observed with adequate spatial resolution.
%
The large frequency of asymmetry reported here suggests that angular 
size measurements and limb darkening studies of AGB stars carried out with 
2-telescope optical long baseline interferometers should be interpreted 
with caution.
We have initiated a systematic mapping program, namely, `The Mira Imaging 
Project' funded by National Science Foundation (NSF) at the IOTA, ISI and 
VLBA interferometers
to connect the asymmetry in space and time, and pin-point the mechanism(s) 
responsible for observed asymmetry. 

\acknowledgments
This work was performed in part under contract with the Jet Propulsion 
Laboratory (JPL) through a Michelson Postdoctoral Fellowship to Ragland, 
funded by NASA as an element of the Navigator (planet finder) Program. 
JPL is managed for NASA by the California Institute of Technology. 
We acknowledge support from the NSF through research grants AST-0138303
and AST-0456047. The IOTA is principally supported by the Smithsonian 
Astrophysical Observatory and the Univ. of Massachusetts. The National 
Radio Astronomy Observatory (NRAO) is operated by Associated 
Universities Inc., under cooperative agreement with the NSF. 
We thank the referee for constructively critical comments that have helped us
to significantly improve the paper. This 
research has made use of NASA's Astrophysics Data System Bibliographic
Services and AAVSO's database for light curves.


\begin{thebibliography}{}
\bibitem[Allen(1984)]{1984Ap&SS..99..101A} Allen, D.~A.\ 1984, \apss, 99, 101


\bibitem[Baldwin et al.(1996)]{Baldwin96} Baldwin, J. E., Beckett, M. G., Boysen, R. C., Burns, D., Buscher, D. F. et al. 1996, A\&A, 306, 13L 

\bibitem[Belczy{\' n}ski et al.(2000)]{2000A&AS..146..407B} Belczy{\' n}ski, K., Miko{\l}ajewska, J., Munari, U., Ivison, R.~J., \& Friedjung, M.\ 2000, \aaps, 146, 407


\bibitem[Berger et al.(2004)]{Berger04} Berger, J.-P., Haguenauer, P., 
Kern, P.,  Rousselet-Perraut, K., Malbet, F.,   et al. 2004, 
Interferometry for Optical Astronomy II, Traub, W.A., Ed., Proc. SPIE 4838, 
1099

\bibitem[Blackman et al.(2001)]{2001Natur.409..485B} Blackman, E.~G., Frank, A., Markiel, J.~A., Thomas, J.~H., \& Van Horn, H.~M.\ 2001, \nat, 409, 485

\bibitem[Boothroyd \& Sackmann(1988)]{1988ApJ...328..632B} Boothroyd, A.~I.~\& Sackmann, I.-J.\ 1988, \apj, 328, 632

\bibitem[Bohme (1978)]{Bohme78} Bohme, D.D., 1978 AN, 299, 243
 
\bibitem[Born \& Wolf(1980)]{Born80} Born, M. \& Wolf, E. 1980, Principles of
Optics, 6 edn. (Oxford, UK: Pergamon Press), Section 10.3.1


\bibitem[Bowen \& Willson(1991)]{Bowen91} Bowen, G. H. \& Willson, L. A. 
1991, ApJ, 375L, 53

\bibitem[Brocksopp et al.(2004)]{Brocksopp04} Brocksopp, C., 
Sokoloski, J. L., Kaiser, C., Richards, A. M., Muxlow, T. W. B. et al. 
2004, MNRAS, 347, 430
%

\bibitem[Chagnon et al.(2002)]{Chagnon02} Chagnon, G., Mennesson, B., 
Perrin, G., Coud\'e du Foresto, V., Salom\'e, P. et al. 2002,AJ, 124, 2821

\bibitem[Cotton et al.(2004)]{Cotton04} Cotton, W. D., Mennesson, B., 
Diamond, P. J., Perrin, G., Coud\'e du Foresto, V. et al.
2004, A\&A, 414, 275

\bibitem[Coud\'e du Foresto, Ridgway \& Mariotti(1997)]{Foresto97} 
Coud\'e du Foresto, V., Ridgway, S. \& Mariotti, J.-M. 1997, 
A\&AS, 121, 379

\bibitem[Diamond \& Kemball(2003)]{Diamond03} Diamond, P.J. \&
Kemball, A.J., 2003, ApJ, 599, 1372

\bibitem[Diamond et al.(1994)]{Diamond94} Diamond, P.J.,
Kemball, A.J., Junor, W., Zensus, A., Benson, J. et al. 1994, 
ApJ, 430, L61

\bibitem[Di Benedetto (1993)]{DiBen93} Di Benedetto, G.P. 1993, A\&A, 270, 315

\bibitem[Dyck, van Belle \& Benson (1996)]{Dyck96} Dyck, H. M., van Belle, G. T. \& Benson, J. A. 
1996, 112, 294

\bibitem[Dyck et al. (1996)]{Dyck96B} Dyck, H. M., Benson, J. A., van Belle, G. T. \& Ridgway, S. T., 1996, AJ, 111, 1705

\bibitem[Dyck, van Belle \& Thompson (1998)]{Dyck98} Dyck, H. M., van Belle, G. T. \& Thompson, R.R. 
1998, 116, 981

\bibitem[Frogel, Persson, \& Cohen(1981)]{1981ApJ...246..842F} Frogel, J.~A., Persson, S.~E., \& Cohen, J.~G.\ 1981, \apj, 246, 842

\bibitem[Greenhill et al.(1995)]{Green95} Greenhill, L. J., Colomer, F., Moran, J. M., Backer, D. C., Danchi, W. C. et al. 1995, ApJ, 449, 365  


\bibitem[Haniff et al.(1992)]{Haniff92} Haniff, C. A., Ghez, A. M., 
Gorham, P. W., Kulkarni, S. R., Matthews, K. et al. 1992, AJ, 103, 1662

\bibitem[Hinkle, Scharlach, \& Hall(1984)]{1984ApJS...56....1H} Hinkle, K.~H., Scharlach, W.~W.~G., \& Hall, D.~N.~B.\ 1984, \apjs, 56, 1

\bibitem[Hofmann et al.(2000)]{Hofmann00} Hofmann, K.-H.,  Balega, Y., 
Scholz, M. \& Weigelt, G. 2000, A\&A, 353, 1016

\bibitem[Humphreys et al.(2002)]{2002A&A...386..256H} Humphreys, 
E.~M.~L., Gray, M.~D., Yates, J.~A., Field, D., 
Bowen, G.~H., \& Diamond, P.~J.\ 2002, \aap, 386, 
256

\bibitem[Hutter et al.(1989)]{Hutter89} Hutter, D. J., Johnston, K. J., Mozurkewich, D., Simon, R. S., Colavita, M. M. et al. 1989, ApJ, 340, 1103


\bibitem[Imai et al.(2002)]{Imai02} Imai, H., Obara, K., 
Diamond, P.J., Omodaka, T. \& Sasao, T. 2002, Nature, 417, 829

\bibitem[Jennison(1958)]{Jennison58} Jennison, R.C. 1958, MNRAS, 118, 276


\bibitem[Knapp \& Morris(1985)]{Knapp85} Knapp, G.~R.~\& Morris, M.\ 1985, \apj, 292, 640

\bibitem[Karovska et al.(1991)]{Karovska91} Karovska, M., Nisenson, P., 
Papaliolios, C. \& Boyle, R.P. 1991, ApJ, 374, L51

\bibitem[Karovska et al.(1997)]{Karovska97} Karovska, M., Hack, W., 
Raymond, J. \& Guinan, E. 1997, ApJ, 482, L175

\bibitem[Kellogg, Pedelty \& Lyon(2001)]{Kellogg01} Kellogg, E., 
Pedelty, J. A. \& Lyon, R. G. 2001, ApJ, 563L, 151
%

\bibitem[Kraus \& Schloerb(2004)]{Kraus04} Kraus, S. \& Schloerb, F.P. 2004, 
New Frontiers in Stellar Interferometry, Traub, W.A., Ed., Proc. SPIE 5491, 56

\bibitem[Lattanzi et al.(1997)]{Lattanzi97} Lattanzi, M.G., Munari, U., 
Whitelock, P.A. \& Feast, M.W. 1997, ApJ, 485, 328

\bibitem[Luthardt(1992)]{1992RvMA....5...38L} Luthardt, R.\ 1992, Reviews of Modern Astronomy, 5, 38

\bibitem[Lopez et al.(1997)]{Lopez97} Lopez, B., Danchi, W. C., Bester, M.,
Hale, D. D. S., Lipman, E. A. et al. 1997, ApJ, 488, 807

\bibitem[Mastrodemos \& Morris(1998)]{1998ApJ...497..303M} Mastrodemos, N.~\& Morris, M.\ 1998, \apj, 497, 303

\bibitem[Mastrodemos \& Morris(1999)]{1999ApJ...523..357M} Mastrodemos, N.~\& Morris, M.\ 1999, \apj, 523, 357

\bibitem[Mennesson et al. (2002)]{Mennesson02} Mennesson, B., Perrin, G., Chagnon, G., 
Foresto, V. Coude du, Ridgway, S. et al. 2002, ApJ, 579, 446

\bibitem[Merrill(1956)]{Merrill56} Merrill, P.W., 1956, PASP, 68, 162
  
\bibitem[Millan-Gabet et al.(2005)]{Millan05} Millan-Gabet, R., Pedretti, E., Monnier, J. D., Schloerb, F. P., 
 Traub, W. A. et al. 2005, ApJ, 620, 961
%

\bibitem[Monnier, Tuthill \& Danchi (2000)]{Monnier00} Monnier, J. D., Tuthill, P. G., Danchi, W. C. 2000, ApJ, 545, 957

\bibitem[Monnier et al.(2004b)]{Monnier04b} Monnier, J. D., Traub, W. A., 
Schloerb, F. P., Millan-Gabet, R., Berger, J.-P. et al. 2004a, 
ApJ, 602L, 57

\bibitem[Monnier et al.(2004a)]{Monnier04a} Monnier, J. D., 
Millan-Gabet, R., Tuthill, P. G., Traub, W. A., Carleton, N. P. et al. 
2004b, ApJ, 605, 436

\bibitem[Mozurkewich \& Armstrong(1992)]{Mozurkewich92} Mozurkewich, D. \& 
Armstrong, J.T. 1992, {\it High-Resolution Imaging by Interferometry II}, 
Beckers, J.M. and Merkle, F., eds. ESO Conference and 
Workshop Proceedings, {\bf 39}, 801

\bibitem[Olofsson et al.(1990)]{1990A&A...230L..13O} Olofsson, H., 
Carlstrom, U., Eriksson, K., Gustafsson, B. \& Willson, L.~A.\ 1990, \aap, 
230, L13

\bibitem[Ostlie \& Cox(1986)]{Ostlie86} Ostlie, D.A. \& Cox, A.N. 1986, ApJ, 311, 864

\bibitem[Pedretti et al.(2004)]{Pedretti04} Pedretti, E., Millan-Gabet, R.,
Monnier, J. D., Traub, W. A., Carleton, N. P., Berger, J.-P., 
Lacasse, M. G., Schloerb, F. P., Brewer, M. K. 2004, PASP, 116, 377

\bibitem[Pedretti et al.(2005)]{Pedretti05} Pedretti, E. Traub, W.A., Monnier, J.D.,
Millan-Gabet, R.; Carleton, N.P., Schloerb, F.P., Brewer, M.K., Berger, J.-P., 
Lacasse, M.G., Ragland, S. 2005, ApOpt., 44, 5173 

\bibitem[Perrin et al.(2004)]{Perrin04} Perrin et al., Ridgway, S.T., 
Mennesson, B., Cotton, W.D., Woillez, J. et al., 2004, A\&A, 426, 279

\bibitem[Ragland(1996)]{Ragland96} Ragland, S., 1996, Ph.D. thesis,
 Physical Research Laboratory, India

\bibitem[Ragland et al.(2004)]{Ragland04} Ragland, S., Traub, W.A., 
Berger, J.-P.,  Millan-Gabet, R., Monnier, J.D. et al. 2004, New Frontiers 
in Stellar Interferometry, Traub, W.A., Ed., Proc. SPIE 5491, 1390
 

\bibitem[Reid \& Goldston(2002)] {Reid02}  Reid, M. J., Goldston, J. E., 
2002, ApJ, 568, 931 

\bibitem[Richichi \& Calamai (1991)]{Richichi93} Richichi, A. \& Calamai, G. 1993,
A\&A, 399, 275
  
\bibitem[Richichi, Lisi, Calamai (1991)]{Richichi91} Richichi, A., Lisi, F. \& Calamai, G. 1991,
A\&A, 241, 131

\bibitem[Richichi et al.(1995)]{Richichi95} Richichi, A., Chandrasekhar, T.,
Lisi, F., Howell, R. R., Meyer, C. et al.
1995, A\&A, 301, 439

\bibitem[Reimers \& Cassatella(1985)]{1985ApJ...297..275R} Reimers, D.~\& Cassatella, A.\ 1985, \apj, 297, 275

\bibitem[Sahai et al.(2003)]{Sahai03} Sahai, R., Morris, M., Knapp, G. R.,
Young, K. \& Barnbaum, C., 2003, Nature, 426, 261
%

\bibitem[Samus et al.(2004)]{Samus04} Samus N.N., Durlevich O.V. et al. 
2004, Combined General Catalog of Variable Stars (GCVS4.2), Institute of 
Astronomy of Russian Academy of Sciences and Sternberg State Astronomical 
Institute of the Moscow State University

\bibitem[Schwarzschild(1975)]{1975ApJ...195..137S} Schwarzschild, M.\ 1975, \apj, 195, 137

%

\bibitem[Soker \& Zoabi(2002)]{2002MNRAS.329..204S} Soker, N.~\& Zoabi, E.\ 2002, \mnras, 329, 204

\bibitem[Sokoloski \& Kenyon(2003)]{Sokoloski03} Sokoloski, J. L. \& 
Kenyon, S. J. 2003, ApJ, 584, 1021

\bibitem[Soria-Ruiz et al.(2004)]{Soria04} Soria-Ruiz, R., Alcolea, J., 
Colomer, F., Bujarrabal, V., Desmurs, J.-F. et al., 2004, A\&A, 426, 131



\bibitem[Struck, Cohanim, \& Willson(2004)]{2004MNRAS.347..173S} Struck, C., Cohanim, B.~E. \& Willson, L.~A.\ 2004, \mnras, 347, 173

\bibitem[Su(2004)]{2004apnw.conf..247S} Su, K.~Y.~L.\ 2004, ASP
Conf.~Ser.~313: Asymmetrical Planetary Nebulae III: Winds, Structure and the Thunderbird, 247

\bibitem[Sudol et al.(2002)]{Sudol02} Sudol, J. J., Benson, J. A., Dyck, H. M., Scholz, M., 
2002, AJ, 124, 3370

\bibitem[Templeton, Mattei \& Willson (2004)]{Templeton04} Templeton, M.R.,
Mattei, J.A. \& Willson, L.A. 2004, private communication

\bibitem[Tuthill et al.(1997)]{Tuthill97} Tuthill, P. G., 
Haniff, C. A. \& Baldwin, J. E. 1997, MNRAS, 285, 529

\bibitem[Tuthill et al.(1999)]{Tuthill99} Tuthill, P. G., 
Haniff, C. A. \& Baldwin, J. E. 1999, MNRAS, 306, 353

\bibitem[Tuthill et al.(2000)]{Tuthill00} Tuthill, P. G., Danchi, W. C., 
Hale, D. S., Monnier, J. D. \& Townes, C. H. 2000, ApJ, 534, 907

\bibitem[Traub et al.(2004)]{Traub04} Traub, W.A., Berger, J.-P., 
Brewer, M.K., Carleton, N.P.,  Kern, P., et al. 2004, 
New Frontiers in Stellar Interferometry, Traub, W.A., Ed., 
Proc. SPIE 5491, 5491, 482

\bibitem[Thompson et al.(2002)]{Thompson02} Thompson, R.R., 
Creeck-Eakman, M.J. \& Akeson, R.L. 2002, ApJ, 570, 373 

\bibitem[van Belle (1999)]{vanBelle99} van Belle, G. T. 1999, PASP, 111, 1515

\bibitem[van Belle, Thompson \& Creeck-Eakman (2002)]{vanBelle02} van Belle, G. T., Thompson, R. R. \& Creech-Eakman, M. J., 2002, AJ, 124, 1706
 
\bibitem[van Belle et al. (1997)]{vanBelle97} van Belle, G. T., Dyck, H. M., Thompson, R. R., 
Benson, J. A. \& Kannappan, S. J. 1997, AJ, 114, 2150
  
\bibitem[van Belle et al. (1996)]{vanBelle96} van Belle, G. T., Dyck, H. M., Benson, J. A. \& 
Lacasse, M. G., 1996, AJ, 112, 2147

\bibitem[Vassiliadis \& Wood (1992)]{1992PASAu..10...30V} Vassiliadis, E.~\& Wood, P.~R.\ 1992, Proceedings of the Astronomical Society of Australia, 10, 30

\bibitem[Weigelt (1977)] {Weigelt77}  Weigelt, G. P. 1977, Opt. Comm., 
21, 55

\bibitem[Weigelt et al. (1996)]{Weigelt96} Weigelt, G., Balega, Y., 
Hofmann, K.-H. \& Scholz, M. 1996, A\&A, 316, L21

\bibitem[Weiner et al.(2006)]{Weiner06} Weiner, J., Tatebe, K., Hale, D. D. S., 
Townes, C. H., Monnier, J. D. et al. 2006, ApJ, 636, 1067

\bibitem[Wesselink, Paranya \& de Vorkin (1972)]{Wesselink72} Wesselink, A. J., Paranya, K. \& de Vorkin, K., 1972, A\&AS, 7, 257

\bibitem[Whitelock (1987)]{1987PASP...99..573W} Whitelock, P.~A.\ 1987, \pasp, 99, 573


\bibitem[Willson (2000)]{2000ARA&A..38..573W} Willson, L.~A.\ 2000, \araa, 38, 573

\bibitem[Willson \& Kim (2004)]{2004apnw.conf..394W} Willson, L.~A.~\& Kim, A.\ 2004, ASP Conf.~Ser.~313: Asymmetrical Planetary Nebulae III: Winds, Structure and the Thunderbird, 394

\bibitem[Wilson et al.(1992)]{Wilson92} Wilson, R.W., Baldwin, J.E., 
Buscher, D.F. \& Warner, P.J. 1992, MNRAS, 257, 369

\bibitem[Wittkowski et al.(1998)]{Wittkowski98} Wittkowski, M., Langer, N. \& 
Weigelt, G. 1998, A\&A, 340L, 39

\bibitem[Wood \& Karovska(2004)]{2004ApJ...601..502W} Wood, B.~E.~\& Karovska, M.\ 2004, \apj, 601, 502

\bibitem[Wood \& Zarro(1981)]{1981ApJ...247..247W} Wood, P.~R.~\& Zarro, D.~M.\ 1981, \apj, 247, 247

\bibitem[Zuckerman \& Aller(1986)]{Zuckerman86} Zuckerman, B. \&
Aller, L. H. 1986, ApJ, 301, 772
\end{thebibliography}
\end{document}